\documentclass[nofootinbib,aps]{revtex4}%
\usepackage{amsmath}
\usepackage{amsfonts}
\usepackage{amssymb}
\usepackage{graphicx}%
\begin{document}
\title[stability of flat space]{Black hole pair creation and the stability of flat space}
\author{Remo Garattini}
\email{Remo.Garattini@unibg.it}
\affiliation{Universit\`{a} degli Studi di Bergamo, Facolt\`{a} di Ingegneria,}
\affiliation{Viale Marconi 5, 24044 Dalmine (Bergamo) Italy}
\affiliation{and I.N.F.N. - sezione di Milano, Milan, Italy.}

\begin{abstract}
We extend the Gross-Perry-Yaffe approach of hot flat space instability to
Minkowski space. This is done by a saddle point approximation of the partition
function in a Schwarzschild wormhole background which is coincident with an
eternal black hole. The appearance of an instability in the whole manifold is
here interpreted as a black hole pair creation.

\end{abstract}
\maketitle

\section{Introduction}

\label{p0}The issue of stability with respect to quantum fluctuations in the
path integral approach to quantum gravity is a fundamental problem as
important as quantum gravity itself. Due to the attractive nature of the
gravitational field, we expect to find a lot of unstable physical
configurations. Therefore one might worry about the stability of the ground
state of quantum gravity. One generally accepts to regard Minkowski space as
the ground state, or vacuum, of quantum gravity. Classical small perturbations
about this vacuum are stable. This is guaranteed by Positivity Theorems of
Schoen and Yau\cite{SchoenYau,SchoenYau1}and Witten\cite{Witten}. However one
might find that flat space is quantum mechanically unstable. This problem was
investigated for the first time by Gross, Perry and Yaffe (GPY)\cite{GPY} in
the Euclidean path integral context. They discovered that hot flat space is
unstable and the loss of stability was interpreted as a spontaneous nucleation
of a black hole with an inverse temperature $\beta=8\pi MG$, where $M$ is the
black hole mass and $G$ is the Newton constant. In the semiclassical
approximation to the Euclidean path integral%
\begin{equation}
Z=\int\mathcal{D}g_{\mu\nu}\exp\left(  -I_{g}\left[  g_{\mu\nu}\right]
\right)  ,\label{p01}%
\end{equation}
the decay probability per unit volume and time $\Gamma$ is defined as%
\begin{equation}
\Gamma=A\exp\left(  -I_{cl}\right)  =\exp\left(  -I_{g}\left[  \bar{g}_{\mu
\nu}\right]  \right)  \int\mathcal{D}h_{\mu\nu}\exp\left(  -I_{g}^{\left(
2\right)  }\left[  h_{\mu\nu}\right]  \right)  ,\label{p02}%
\end{equation}
where the gravitational field $g_{\mu\nu}$ has been separated into a
background field $\bar{g}_{\mu\nu}$ and a quantum fluctuation $h_{\mu\nu}$%
\begin{equation}
g_{\mu\nu}=\bar{g}_{\mu\nu}+h_{\mu\nu},
\end{equation}
$A$ is the prefactor coming from the saddle point evaluation of $Z$ and
$I_{cl}$ is the classical part of the action. If a single negative eigenvalue
appears in the prefactor $A$, it means that the related bounce shifts the
energy of the false ground state\cite{Coleman}. In the Schwarzschild case GPY
discovered one single negative mode $\lambda_{neg}=-.19/\left(  MG\right)
^{2}$. Allen reconsidered the effect of finite boundaries on the appearance of
a negative mode showing that, if the box enclosing the black hole is little
enough the instability disappears\cite{Allen}. Since the pioneering paper of
Gross, Perry and Yaffe, many other spacetime configurations have been
investigated in different contexts. In particular, the de Sitter case
involving a positive cosmological constant has been examined in the
inflationary context by Bousso and Hawking\cite{BoHaw}. In the same context of
the de Sitter background, Ginsparg and Perry\cite{GP}, Young\cite{Young} and
more recently Volkov and Wipf\cite{VW} have computed the partition function to
one loop. On the other hand, the Anti-de Sitter case involving a negative
cosmological constant has been discussed by Hawking and Page\cite{HawPage} and
subsequently by Prestidge\cite{Prestidge}. The parallel problem in a higher
dimensional context of the GPY-instability was firstly taken under examination
by Witten\cite{Witten1}, who examined the stability of Kaluza-Klein theories.
After a decade Gregory and Laflamme reconsidered the problem of the
gravitational stability in the context of branes\cite{GregoryLaflamme}. This
opened an interest in string theory where the stability of black branes was
widely examined by different authors\cite{authors}. It is interesting to note
that the only case of instability involving not a nucleation of a single black
hole but a pair, is the de Sitter case. Moreover a deep difference appears
between this case and the Anti-de Sitter or Schwarzschild case: temperatures
before and after pair creation are different, at least in the neutral de
Sitter black hole pair creation. Motivated by this result, we would like to
reconsider the stability of flat space without involving temperatures as in
the GPY analysis, but looking at a neutral black hole pair creation mediated
by a wormhole of the Schwarzschild type. Recall that in the GPY instability is
not possible to reach a vanishing temperature, because $T=1/\beta=1/\left(
8\pi MG\right)  $. Therefore to investigate the stability at zero temperature,
we need to consider a Schwinger-like process of black hole pair
creation\footnote{A recent detailed historical overview about the black hole
pair creation process can be found in Ref.\cite{DiasLemos}.}. The
Schwinger-like process has been investigated in a series of papers in a
variational approach based on a Hamiltonian formulation of the Einstein
gravity\cite{Remo}. In this paper, we would like to apply the GPY\ analysis to
such a problem. The rest of the paper is structured as follows, in section
\ref{p2} we write the gravitational action for the Schwarzschild wormhole, in
section \ref{p3} we follow the GPY procedure of separating TT modes, in
section \ref{p4} analyze the eigenvalue equation. We summarize and conclude in
section \ref{p5}. Units in which $\hbar=c=k=1$ are used throughout the paper.

\section{Gravitational Action with boundaries for the Schwarzschild wormhole}

\label{p2}
\begin{figure}
[th]
\begin{center}
\includegraphics[
height=2.2864in,
width=3.2237in
]%
{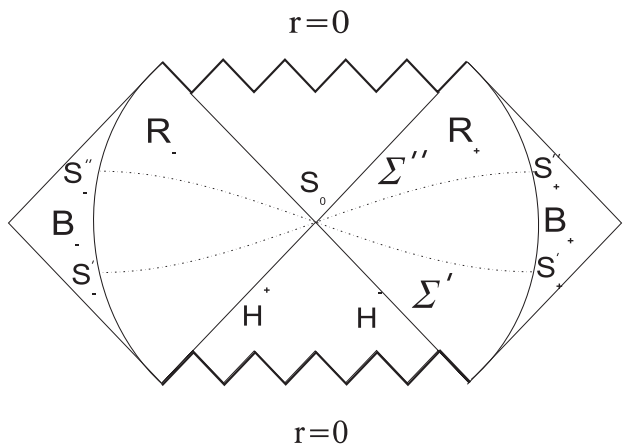}%
\caption{This is a Penrose diagram for the Schwarzschild wormhole.}%
\label{penrose}%
\end{center}
\end{figure}

\begin{figure}
[th]
\begin{center}
\includegraphics[
height=2.0274in,
width=2.3952in
]%
{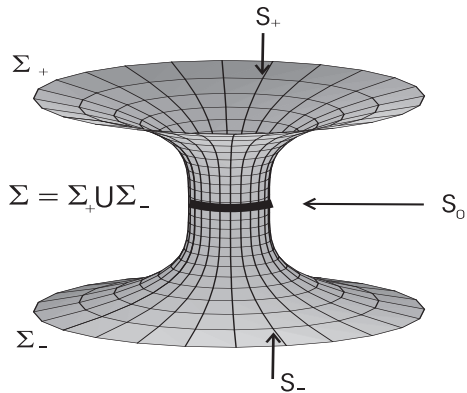}%
\caption{Wormhole representation for the Schwarzschild metric. The bifurcation
surface $S_{0}$ divides the hypersurfaces $\Sigma_{+}$ and $\Sigma_{-}$.}%
\label{wormhole}%
\end{center}
\end{figure}

Before considering the perturbations on the Schwarzschild wormhole,
we describe some properties of the Schwarzschild-Kruskal line element%
\begin{equation}
ds^{2}=-\frac{32\left(  MG\right)  ^{3}}{r}\exp\left(  -\frac{r}{2MG}\right)
\,dU\,dV+r^{2}d\Omega^{2}. \label{Kruskal}%
\end{equation}
$M$ represents the Arnowitt-Deser-Misner (ADM) mass of the wormhole\cite{ADM},
$d\Omega^{2}$ is the line element of the unit two-sphere, and the radial
coordinate is regarded as a function of the ingoing and outgoing null
coordinates $(U,V)$ as
\begin{equation}
\left(  1-\frac{r}{2MG}\right)  \exp\left(  -\frac{r}{2MG}\right)  =UV\ ,
\end{equation}
while the time coordinate regarded as a function of the same coordinates
$\left(  U,V\right)  $ is defined as%
\begin{equation}
t=2MG\ln\left\vert \frac{-V}{U}\right\vert .
\end{equation}
The Schwarzschild-Kruskal spacetime is the union of four regions (wedges)
$R_{+}$, $R_{-}$, $T_{+}$, and $T_{-}$. The regions $R_{+}$ and $R_{-}$ are
asymptotically flat . In the $R_{+}$ region, the Kruskal coordinates $(U,V)$
satisfy $U<0$, $V>0$, while in $R_{-}$, $U>0$, $V<0$. An important property of
the line element $\left(  \ref{Kruskal}\right)  $ is its invariance with
respect to the discrete symmetries%
\begin{equation}
\mathbf{I}:U\rightarrow-U;\,V\rightarrow-V,\qquad\mathbf{T}:U\rightarrow
-V;\,V\rightarrow-U\ . \label{discrete}%
\end{equation}
This means that we expect the physics be symmetric with respect to the
bifurcation surface ${S}_{0}$ defined as the intersection of $H^{+}$ and
$H^{-}$, which are the future $\left(  H^{+}\right)  $ and past $\left(
H^{-}\right)  $ horizons, respectively. In Fig.$\left(  \ref{penrose}\right)
$, we show the corresponding Penrose diagram%
We denote with $\Sigma_{t}$ the $t=constant$ hypersurface. Any of such
hypersurfaces is invariant under $T$-reflections and has Einstein-Rosen bridge
topology $R^{1}\times S^{2}$, represented in Fig.$\left(  \ref{wormhole}%
\right)  $, whose intrinsic geometry and time derivatives are chosen to
satisfy the gravitational constraint equations%
We denote by ${\Sigma}_{\pm}$ the parts of ${\Sigma}$ lying in $R_{\pm}$
respectively. The line element $\left(  \ref{Kruskal}\right)  $ restricted to
the surface ${\Sigma}$ reads
\begin{equation}
dl^{2}=dy^{2}+r^{2}(y)d\Omega^{2}\ ,
\end{equation}
where the quantity $y$, defined by
\begin{equation}
dy=\pm\frac{dr}{\sqrt{1-2MG/r}}, \label{propdist}%
\end{equation}
represents the proper geodesic distance from the \textquotedblleft
throat\textquotedblright\ of the bridge located at $r=r(y=0)=r_{+}=2MG$. We
choose $y$ to be positive in ${\Sigma}_{+}$ and negative in ${\Sigma}_{-}$.
The set $(t,y,\theta,\phi)$ can be used as canonical coordinates everywhere in
$R_{\pm}$ outside the bifurcation surface ${S}_{0}$. These coordinates are
right-oriented in $R_{+}$ and left-oriented in $R_{-}$. The configuration we
wish to examine is described by boundary surfaces located in different regions
with two different slices ${\Sigma}^{\prime}$ $\equiv\Sigma_{t^{\prime}}$ and
${\Sigma}^{\prime\prime}\equiv\Sigma_{t^{\prime\prime}}$ intersecting at the
same two-dimensional bifurcation sphere ${S}_{0}$. We call this sequence of
slices ${\Sigma}_{t}$ (defined by the equation $t=\mathrm{constant}$, with
$t^{\prime}\leq t\leq t^{\prime\prime}$) a \textquotedblleft tilted
foliation\textquotedblright\cite{FroMar}. For a \textquotedblleft tilted
foliation\textquotedblright\ the spacetime domain $M$ lying between ${\Sigma
}^{\prime}$ and ${\Sigma}^{\prime\prime}$ consists of two wedges
$\mathcal{M}_{+}$ and $\mathcal{M}_{-}$ located in the right ($R_{+}$) and
left ($R_{-}$) sectors of the Kruskal diagram. The region $\mathcal{M}%
=\mathcal{M}_{+}\cup\mathcal{M}_{-}$ is bounded by $\Sigma^{\prime}$ and
$\Sigma^{\prime\prime}$ and by a three-dimensional timelike boundary $B$ that
consists of two disconnected parts $B_{+}$ and $B_{-}$. For a general eternal
black hole geometry the boundaries $B_{+}$ and $B_{-}$ are located in
$\mathcal{M}_{+}$ and $\mathcal{M}_{-}$, respectively. We define ${S_{\pm}%
^{2}=}\Sigma_{\pm}\cap{B_{\pm}}$. The topology of the slices $\Sigma$ is
therefore $I_{\Sigma}\times S^{2}$, where $I_{\Sigma}$ is a finite spacelike
distance, while the topology of ${B_{\pm}}$ is $I_{t}\times S_{\pm}^{2}$,
where $I_{t}$ is a finite timelike distance. In Schwarzschild coordinates the
line element of Eq.$\left(  \ref{Kruskal}\right)  $ can be written as%
\begin{equation}
ds^{2}=-N^{2}dt^{2}+dy^{2}+r^{2}\left(  y\right)  d\Omega^{2}, \label{a1a}%
\end{equation}
where $y$ is the proper radial distance from the throat defined in Eq.$\left(
\ref{propdist}\right)  $. The coordinate $y$ covers the entire range $\left(
-\infty,+\infty\right)  $. $N$ is the corresponding lapse function written in
terms of the proper distance from the throat
\begin{equation}
N^{2}=1-\frac{2MG}{r\left(  y\right)  }. \label{lapse}%
\end{equation}
We define the four-velocity vector with $u_{\mu}=-N\partial_{\mu}t$, while the
timelike unit vector normal to the hypersurface $\Sigma_{t\,}$ is $u^{\mu}$.
This is chosen to be future oriented in $\mathcal{M}$ and normalized by the
condition $u\cdot u=-1$. The lapse $N$ is positive at $\Sigma_{+}$, negative
at $\Sigma_{-}$ and vanishing at the bifurcation surface. The spacelike normal
$n^{\mu}$ to the three-dimensional boundaries $B{_{\pm}}$, is defined to be
outward pointing at $B{_{+}}$, inward pointing at $B{_{-}}$, and normalized so
that $n\cdot n=1$ with the further condition $\left.  \left(  u\cdot n\right)
\right\vert _{{B_{\pm}}}=0$. Following Ref.\cite{FroMar}, greek indices are
used for tensors in $\mathcal{M}$ while latin indices are used for tensors
defined in either $\Sigma$ or $B{_{\pm}}$. The metric and extrinsic curvature
of $\Sigma$ as a surface embedded in $\mathcal{M}$ are denoted by $h_{ij}$ and
$K_{ij}=-h_{i}^{k}\nabla_{k}u_{j}$, respectively, while the metric and
extrinsic curvature of the boundaries ${B_{\pm}}$ as surfaces embedded in
$\mathcal{M}$ are $\gamma_{ij}$ and $\Theta_{ij}=-\gamma_{i}^{k}\nabla
_{k}n_{j}$. Covariant differentiation with respect to the metric $g_{\mu\nu}$
and $h_{ij}$ is denoted by $\nabla$ and $D$, respectively. The induced metric
and extrinsic curvature of the boundaries ${S_{\pm}^{2}}$ as surfaces embedded
on $\Sigma$ are denoted by $\sigma_{ab}$ and $k_{ab}=-\sigma_{a}^{k}D_{k}%
n_{b}$, respectively, $\left(  a,b=2,3\right)  $. Explicitly, the metric
tensors for the different surfaces are%
\begin{equation}
h_{\mu\nu}=g_{\mu\nu}+u_{\mu}u_{\nu},\qquad\gamma_{\mu\nu}=g_{\mu\nu}-n_{\mu
}n_{\nu},\qquad\sigma_{\mu\nu}=g_{\mu\nu}+u_{\mu}u_{\nu}-n_{\mu}n_{\nu}.
\end{equation}
The determinants of the metric tensors are related by
\begin{equation}%
\begin{array}
[c]{c}%
\sqrt{-g}=\left\vert N\right\vert \sqrt{h},\\
\sqrt{-\gamma}=\left\vert N\right\vert \sqrt{\sigma}.
\end{array}
\end{equation}
The covariant form for the gravitational action generated by a tilted
foliation with fixed three-dimensional boundary of $\mathcal{M}$
is\footnote{The complete action should be considered with additional boundary
terms of the type
\begin{equation}
\frac{1}{\kappa}\int\limits_{t^{\prime},\Sigma_{\pm}}^{t^{\prime\prime}}%
d^{3}x\,\sqrt{h}K_{\pm}.
\end{equation}
Nevertheless, since we will look at the Euclideanized version of the action
with periodic boundary conditions in the Euclidean time, the previous boundary
term disappears.}
\[
S=\frac{1}{2\kappa}\int_{\mathcal{M}_{+}}d^{4}x\,\sqrt{-g}\,\Re-\frac
{1}{\kappa}\int_{B_{+}}d^{3}x\,\sqrt{-\gamma}\left(  \,\Theta-\,\Theta
_{0}\right)  _{+}%
\]%
\begin{equation}
-\frac{1}{2\kappa}\int_{\mathcal{M}_{-}}d^{4}x\,\sqrt{-g}\,\Re-\frac{1}%
{\kappa}\int_{B_{-}}d^{3}x\,\sqrt{-\gamma}\left(  \,\Theta-\,\Theta
_{0}\right)  _{-}, \label{a3}%
\end{equation}
where $\Re$ denotes the four-dimensional scalar curvature, $\kappa\equiv8\pi
G$. The integrations are taken over the coordinates $x^{\mu}$ which have the
same orientation as the canonical coordinates $(t,y,\theta,\phi)$ of the
tilted foliation. The differing signs in $\mathcal{M}_{+}$ and $\mathcal{M}%
_{-}$ reflect the fact that the coordinates have different time orientations
in $\mathcal{M}_{+}$ and $\mathcal{M}_{-}$. The subtraction term
\begin{equation}
S_{0\left(  \pm\right)  }={\frac{1}{\kappa}}\int_{{B_{\pm}}}d^{3}%
x\,\sqrt{-\gamma}\,\Theta_{0}%
\end{equation}
is the extrinsic curvature evaluated on the reference space, which in this
case has been chosen to be flat. The effect of $S_{0\left(  \pm\right)  }$ is
to normalize the energy $E$ to zero for the Schwarzschild wormhole with $M=0$.
In the next section we will carefully examine the contribution of the volume
terms of the action. Here, we restrict the evaluation on the boundaries
$B_{\pm}$ obtaining%
\begin{equation}
S=-\frac{1}{\kappa}\int_{B_{+}}d^{3}x\,\sqrt{-\gamma}\left(  \,\Theta
-\,\Theta_{0}\right)  _{+}-\frac{1}{\kappa}\int_{B_{-}}d^{3}x\,\sqrt{-\gamma
}\left(  \,\Theta-\,\Theta_{0}\right)  _{-}. \label{a4}%
\end{equation}
The trace of the extrinsic curvature is%
\begin{equation}
\Theta_{|_{\pm}}=-\nabla_{k}n^{k}=-\frac{1}{\sqrt{-g}}\partial_{k}\left(
\sqrt{-g}g^{kj}n_{j}\right)  =-\frac{1}{\left\vert N\right\vert r^{2}}%
\partial_{y}\left(  \left\vert N\right\vert r^{2}\right)  . \label{trace}%
\end{equation}
With the help of Eqs.$\left(  \ref{propdist}\right)  $ and $\left(
\ref{lapse}\right)  $, we can write%
\begin{equation}
\frac{\partial\left\vert N\right\vert }{\partial y}=\pm\frac{MG}{r^{2}\left(
y\right)  }\qquad\text{in }\Sigma_{\pm},
\end{equation}
where we have used%
\begin{equation}
\frac{\partial r}{\partial y}=\pm\sqrt{1-\frac{2MG}{r\left(  y\right)  }%
}\qquad\text{in }\Sigma_{\pm}.
\end{equation}
Thus Eq.$\left(  \ref{trace}\right)  $ becomes
\begin{equation}
\Theta_{|_{\pm}}=\mp\frac{1}{\left\vert N\right\vert r^{2}\left(  y\right)
}\left[  2r\left(  y\right)  -3MG\right]  ,
\end{equation}
while the subtraction term can be defined by
\begin{equation}
\Theta_{0|_{\pm}}=\lim_{M\rightarrow0}\Theta_{|_{\pm}}=\mp\frac{2}{r}%
\end{equation}
and
\begin{equation}
\left(  \Theta-\,\Theta_{0}\right)  _{\pm}=\mp\frac{1}{\left\vert N\right\vert
r^{2}\left(  y\right)  }\left[  2r\left(  y\right)  -3MG-2\frac{r^{2}\left(
y\right)  }{r}\left\vert N\right\vert \right]  .
\end{equation}
The proper \textquotedblleft\textit{time}\textquotedblright\ length in
Eq.$\left(  \ref{a4}\right)  $ is
\begin{equation}
{\frac{1}{\kappa}}\int_{{B}_{\pm}}dt\left\vert N\right\vert ={\frac{1}{\kappa
}T}\left\vert N\left(  y_{\pm}\right)  \right\vert \qquad\text{in }\Sigma
_{\pm}, \label{a4a}%
\end{equation}
where $y_{\pm}$ is the location of the spatial boundary on $B_{\pm}$ and
$T=t^{\prime\prime}-t^{\prime}$. The substitutions $t\rightarrow-i\tau$ and
$N\rightarrow-iN_{E}$ do not modify the behavior of the boundary action in the
Euclidean context. Thus Eq.$\left(  \ref{a4a}\right)  $ becomes%
\begin{equation}
\beta_{\pm}=T_{\pm}^{-1}=N_{E}\left(  r_{\pm}\right)  \beta^{\ast}%
\qquad\text{in }\Sigma_{\pm},
\end{equation}
where $r_{\pm}$ stands for $r\left(  y_{\pm}\right)  $ and where we have
integrated over the period $\beta^{\ast}=8\pi MG$ obtained by imposing
regularity condition on the horizon. Note that the inverse temperature at
infinity is the same in both wedges. The difference between $\beta_{+}$ and
$\beta_{-}$ essentially arises from the Tolman law involving the different
location of the boundaries. Then the total action contribution is
\begin{equation}
S_{{B}_{\pm}}=\pm{\frac{4\pi}{\kappa}T}\left[  2r-3MG-2rN_{E\pm}\right]
|_{r=r_{\pm}}\longrightarrow I_{{B}_{\pm}}=\mp\left(  8\pi Mr-12\pi
M^{2}G-\beta r/G\right)  |_{r=r_{\pm}}.
\end{equation}
Like in the case of the Hamiltonian discussed in Ref.\cite{FroMar}, the tree
level action for the Schwarzschild wormhole can be cast in the form
\begin{equation}
\frac{1}{\kappa}\left(  I_{+}-I_{-}\right)  =\left(  12\pi M^{2}G-8\pi
Mr_{+}+\beta_{+}r_{+}/G\right)  -\left(  12\pi M^{2}G-8\pi Mr_{-}+\beta
_{-}r_{-}/G\right)  . \label{a5}%
\end{equation}
From Eq.$\left(  \ref{a5}\right)  $, we can compute the energy of the system
at the fixed temperature $\beta_{\pm}$ in each wedge. This is obtained by
regarding Eq.$\left(  \ref{a5}\right)  $ as a function of $r$ and $T$ rather
than of $r$ and $M$. We obtain
\begin{equation}%
\begin{array}
[c]{cc}%
E_{+}=\frac{\partial I_{+}}{\partial\beta_{+}} & =\frac{1}{G}\left[
r_{+}-r_{+}\sqrt{1-2MG/r_{+}}\right] \\
E_{-}=\frac{\partial I_{-}}{\partial\beta_{-}} & =\frac{1}{G}\left[
r_{-}-r_{-}\sqrt{1-2MG/r_{-}}\right]
\end{array}
. \label{a6}%
\end{equation}
Note that each term tends asymptotically to the Arnowitt-Deser-Misner
$(\mathcal{ADM})$ mass\cite{ADM}. Note also that if the boundary location is
different in the respective wedge, implies that $E_{+}\neq E_{-}$.
Nevertheless for the entropy we discover
\begin{equation}%
\begin{array}
[c]{cc}%
S_{+}=\beta_{+}\frac{\partial I_{+}}{\partial\beta_{+}}-I_{+} & =\beta
_{+}E_{+}-I_{+}=4\pi M^{2}G\\
S_{-}=\beta_{-}\frac{\partial I_{-}}{\partial\beta_{-}}-I_{-} & =\beta
_{-}E_{-}-I_{-}=4\pi M^{2}G
\end{array}
, \label{a7}%
\end{equation}
namely it is independent on the wedge. This is a consequence of the definition
we have used for computing the entropy. The same result has been obtained by
Martinez in Ref.\cite{Martinez} in the context of microcanonical approach to
the entropy of an eternal black hole. Note that if one adopts the method used
in Ref.\cite{Martinez} for computing the entropy, it is not immediate to
arrive at the result of Eq.$\left(  \ref{a7}\right)  $. By means of
Eq.$\left(  \ref{a7}\right)  $, Eq.$\left(  \ref{a5}\right)  $ can be
explicitly rewritten as
\begin{equation}
\beta_{+}E_{+}-\beta_{-}E_{-}=8\pi MG\left(  N_{E+}E_{+}-N_{E-}E_{-}\right)
=\beta^{\ast}\left(  M_{+}-M_{-}\right)  .
\end{equation}
Note that $\beta^{\ast}$ is the periodically identified \textquotedblleft
asymptotic Euclidean time\textquotedblright\ which is independent on the
considered wedge. We can recognize three different cases

\begin{enumerate}
\item $M_{+}>M_{-}$. The limit case is when $r_{+}\rightarrow\infty\left(
y_{+}\rightarrow+\infty\right)  $, while $r_{-}=2GM\left(  y_{-}=0\right)  $.
This corresponds to $M_{+}=M_{\mathcal{ADM}}$ and $M_{-}=0$. This situation is
the standard one, where the black hole thermodynamics is investigated in the
wedge $\mathcal{M}_{+}$. Indeed

\item $M_{+}<M_{-}$. The limit case is when $r_{-}\rightarrow\infty\left(
y_{-}\rightarrow-\infty\right)  $, while $r_{+}=2GM\left(  y_{+}=0\right)  $.
This corresponds to $M_{-}=M_{\mathcal{ADM}}$ and $M_{+}=0$. This situation is
``\textit{dual}'' to the previous in the sense that the black hole
thermodynamics is investigated in the wedge $\mathcal{M}_{-}$.

\item $M_{+}=M_{-}$.The left boundary $y_{-}$ and the right boundary $y_{+}$
are symmetric with respect to the bifurcation surface $S_{0}^{2}$, which
implies that $r_{+}=r_{-}$.
\end{enumerate}

Case 3 corresponds to a vanishing action. Nevertheless, it is well known that
a vanishing action with a Euclidean metric describes flat
space\cite{SchoenYau1}. Therefore, in this very particular case we have found
an alternative way to flat space at the price of having a topology change.

\section{Nonconformal Negative Modes}

\label{p3}The measure of stability in this approach is the eigenvalue spectrum
of the nonconformal perturbative modes for the solution. Should there be a
mode with a negative eigenvalue, then the action for this solution is a
saddle-point in its phase space rather than a true minimum. Consequently,
there ought to be a correspondence between the presence of such a negative
mode and the local thermodynamic stability as governed by the heat capacity.
Negative modes arise from the analysis of geometric fluctuations about
classical Euclidean solutions of the Einstein field equations. However, the
analysis to confirm their existence must be performed with care, since the
gauge freedom of the Euclidean action will in general introduce a large number
of non-physical negative modes associated with conformal deformations of the
metric. For pure gravity, the contributions from the conformal and the
nonconformal modes decouple if a suitable gauge is chosen. In the path
integral approach~\cite{Hawking}, the partition function $Z$ is generally
defined as a functional integral over all metrics with some fixed asymptotic
behavior on some manifold $\mathcal{M}$,
\begin{equation}
Z=\int_{\mathcal{M}}D[g]\exp\left(  -iS[g]\right)  .
\end{equation}
This integral is formally defined by an analytic continuation to a Euclidean
section of $\mathcal{M}$ denoted by $\mathcal{\bar{M}}$, where the right and
left wedges of a Lorentzian eternal black hole (wormhole) are mapped into two
complex sectors $\mathcal{\bar{M}}_{+}$ and $\mathcal{\bar{M}}_{-}$ to become
\begin{equation}
Z=\int_{\mathcal{\bar{M}}}D[g]\exp\left(  -I\left[  g\right]  \right)  ,
\label{b0}%
\end{equation}
where the integral is performed over all positive definite metrics $g$. In the
case of pure gravity, the Euclidean action comes from Eq.$\left(
\ref{a3}\right)  $ whose expression is%
\[
I=-{\frac{1}{2\kappa}}\int_{\mathcal{\bar{M}}_{+}}d^{4}x\,\sqrt{g}%
\,R+{\frac{1}{\kappa}}\int_{\partial\mathcal{\bar{M}}{_{+}}}d^{3}%
x\,\sqrt{\gamma}\left(  \,\Theta-\,\Theta_{0}\right)
\]%
\begin{equation}
+{\frac{1}{2\kappa}}\int_{\mathcal{\bar{M}}_{-}}d^{4}x\,\sqrt{g}\,R+{\frac
{1}{\kappa}}\int_{\partial\mathcal{\bar{M}}_{-}}d^{3}x\,\sqrt{\gamma}\left(
\,\Theta-\,\Theta_{0}\right)  . \label{Euaction}%
\end{equation}
This partition function may be approximated using saddle-point techniques, by
Taylor expanding about the known stationary points of the Euclidean action --
the solutions to the Einstein field equations
\begin{equation}
R_{\mu\nu}=0.
\end{equation}
The expansions are performed by writing the metric $g_{ab}$ as
\begin{equation}
g_{\mu\nu}=\bar{g}_{\mu\nu}+h_{\mu\nu}%
\end{equation}
with $h_{\mu\nu}$ treated as a quantum field on the classical fixed background
$\bar{g}_{\mu\nu}$ which vanishes on the boundary $\partial\mathcal{\bar{M}%
}_{\pm}$. Thus the Euclidean action becomes%
\begin{equation}
I[g]=I[\bar{g}]_{+}+I_{2}[h]_{+}-I[\bar{g}]_{-}-I_{2}[h]_{-}+\cdots
\label{expansion}%
\end{equation}
where the linear term $I_{1}$ vanishes precisely because $\bar{g}_{\mu\nu}$ is
a classical solution, $I_{2}$ is quadratic in the field $h_{\mu\nu}$ in each
complex sector, and `$\cdots$' represents terms of higher than quadratic
order. The Taylor expansion of the action in $\mathcal{\bar{M}}=\mathcal{\bar
{M}}_{+}\cup\mathcal{\bar{M}}_{-}$ leads to the following one loop
approximation of the partition function%
\begin{equation}
\log Z=-I\left[  \bar{g}\right]  _{+}+\log\int_{\mathcal{\bar{M}}_{+}}D\left[
h\right]  _{+}\exp\left(  -I_{2}\left[  h\right]  _{+}\right)  +I\left[
\bar{g}\right]  _{-}+\log\int_{\mathcal{\bar{M}}_{-}}D\left[  h\right]
_{-}\exp\left(  I_{2}\left[  h\right]  _{-}\right)  . \label{1loop}%
\end{equation}
It is interesting to note that in this form the gravitational quantum
fluctuations on $\mathcal{\bar{M}}_{+}$ $\left(  \mathcal{\bar{M}}_{-}\right)
$ act as a sort of source field with respect to $\mathcal{\bar{M}}_{-}\left(
\mathcal{\bar{M}}_{+}\right)  $. The quadratic contribution to the action is
straightforward to evaluate, and may be written for arbitrary $h_{\mu\nu}$ in
each wedge, in the form%
\begin{equation}
I_{2}[h]_{\pm}=\frac{1}{4\kappa}\int d^{4}x\sqrt{g}h_{\pm}^{\mu\nu}G_{\mu
\nu\rho\sigma}h_{\pm}^{\rho\sigma}.
\end{equation}
$G$ is a physical gauge invariant spin-2 operator which acts on the transverse
and trace-free part of $h_{\mu\nu}$ and takes the simple form%
\begin{equation}
G_{\mu\nu\rho\sigma}=-g_{\mu\rho}g_{\nu\sigma}\nabla_{\alpha}\nabla^{\alpha
}-2R_{\mu\rho\nu\sigma}.
\end{equation}
In general, $G$ will have some finite number -- generally zero or one -- of
negative eigenvalues, which correspond to the nonconformal negative modes of
the solution. The eigenvalues of $G$ are determined by all solutions to the
elliptic equation
\begin{equation}
G_{\rho\sigma}^{\mu\nu}h_{(n)}^{\rho\sigma}=\lambda_{(n)}h_{(n)}^{\mu\nu}
\label{eigenvalue}%
\end{equation}
where the eigenfunctions $h_{\mu\nu}$ are real, regular, symmetric,
transverse, trace-free, and normalizable tensors. Clearly, should one of the
eigenvalues of $G$ be negative, then the product of all of the eigenvalues
would also be negative. The contribution to $\log Z$ from fluctuations about
the classical solution would then contain an imaginary component, leading to
an instability in the ensemble similar to the type proposed by Gross, Perry
and Yaffe~\cite{GPY}. Naively, then, the one-loop term may be written as
$\frac{1}{2}\log Det(\mu^{-2}G)$ where $\mu$ is a regularization mass, and the
determinant is formally defined as the product of the eigenvalues of $G$.
However, due to the diffeomorphism gauge freedom of the action, $A$ will in
general have a large number of zero eigenvalues, and so this procedure as
stated is ill-defined. The remedy is to add a gauge fixing term $B$ -- such
that the operator $A+B$ has no zero eigenvalues -- and an associated ghost
contribution $C$, to obtain
\begin{equation}
\log Z=-I[g]-\frac{1}{2}\log Det(\mu^{-2}\{A+B\})+\log Det(\mu^{-2}C).
\end{equation}
Such terms may be dealt with by means of generalized zeta functions, as
considered by Gibbons, Hawking, and Perry~\cite{GHP}, and extended to include
a $\Lambda$ term by Hawking~\cite{Hawking}. In order that this be possible,
the terms must be expressed as sums of operators, each with only a finite
number of negative eigenvalues. This may be achieved by writing $A+B$ as
$-F+G$, where $F$ is a scalar operator acting on the trace of $h_{\mu\nu}$.
The ghost term $C$ is a spin-1 operator acting on divergence-free vectors, and
An observation of primary significance is that a gauge may be chosen in which
the operators $F$ and $C$ have no negative eigenvalues. If the background
metric $g_{ab}$ is flat then, in addition, $G$ will be positive-definite, but
for a non-flat background this is not the case.

\section{Variational Approach to the Negative Mode}

\label{p4}Following the treatment given in~\cite{GPY}, and subsequently
in~\cite{Allen} for the Schwarzschild case, it is clear that only spherically
symmetric and $\tau$-independent solutions of Eq.$\left(  \ref{eigenvalue}%
\right)  $ need be considered as candidate nonconformal negative modes. In
static and spherically symmetric backgrounds, modes of higher multipole moment
will necessarily have greater eigenvalues. With this assumption, it is then
straightforward to write down a construction for such solutions to $G$ valid
in a four-dimensional Euclidean wormhole background of the form $\left(
\ref{a1a}\right)  $. Since $G$ acts only on symmetric transverse and
trace-free tensors, then clearly the constructed solutions must exhibit all of
these properties. If the mode $h_{\mu\nu}^{\pm}$ is written in the manifestly
trace-free and symmetric form%
\begin{equation}
\left(  h_{\nu}^{\mu}\right)  ^{\pm}=diag\left(  H_{0}^{\pm}(y),H_{1}^{\pm
}(y),-\frac{1}{2}\left(  H_{0}^{\pm}(y)+H_{1}^{\pm}(y)\right)  ,-\frac{1}%
{2}\left(  H_{0}^{\pm}(y)+H_{1}^{\pm}(y)\right)  \right)  , \label{tracefree}%
\end{equation}
then the final property, $\nabla^{a}h_{\mu\nu}^{\pm}=0$, is guaranteed if it
is further assumed that $H_{0}^{\pm}(y)$ and $H_{1}^{\pm}(y)$ are related
through the first order equation%
\begin{equation}
H_{1,y}^{\pm}+\frac{3}{r}r_{,y}H_{1}^{\pm}+\frac{N_{,y}}{N}H_{1}^{\pm}%
=\frac{N_{,y}}{N}H_{0}^{\pm}-\frac{3}{r}r_{,y}H_{0}^{\pm}. \label{transverse}%
\end{equation}
In Eqs.$\left(  \ref{tracefree}\right)  $ and $\left(  \ref{transverse}%
\right)  $, we have explicitly written the location of the perturbation on
each patch covering one universe. The two patches join at the throat of the
wormhole and the transversality condition can be cast into the form%
\begin{equation}
\left[  \frac{dH_{1}^{\pm}(r)}{dr}+\frac{3}{r}H_{1}^{\pm}(r)+\frac{1}{N}%
\frac{dN}{dr}H_{1}^{\pm}(r)\right]  \frac{dr}{dy}=\left[  \frac{1}{N}\frac
{dN}{dr}H_{0}^{\pm}(r)-\frac{3}{r}H_{0}^{\pm}(r)\right]  \frac{dr}{dy},
\label{transverse1}%
\end{equation}
where it is manifest the independence on the wedge location. This is a
consequence of the discrete symmetry $\left(  \ref{discrete}\right)  $ between
the asymptotically flat wedges. When we apply this symmetry to the the
eigenvalue equation $\left(  \ref{diff_eq}\right)  $ with the help of the
following substitution%
\begin{align*}
N  &  \rightarrow-N\\
r_{,y}  &  \rightarrow-r_{,y},
\end{align*}
we find that Eq.$\left(  \ref{diff_eq}\right)  $ is invariant in form only on
the coordinate $r$. With the ansatz $\left(  \ref{tracefree}\right)  $ and
$\left(  \ref{transverse1}\right)  $, the eigenvalue equation $\left(
\ref{eigenvalue}\right)  $ reduces to a linear second order ordinary
differential equation for the component $H_{1}^{\pm}\left(  y\right)  $ and
eigenvalue $\mu^{2}.$ To further proceed, we insert Eqs.$\left(
\ref{propdist}\right)  $ and $\left(  \ref{lapse}\right)  $ into Eq.$\left(
\ref{eigenvalue}\right)  $. Since we have chosen $y$ and the Euclidean lapse
$N_{E}$ to be positive in $\mathcal{\bar{M}}_{+}$ and negative in
$\mathcal{\bar{M}}_{-}$, then Eq.$\left(  \ref{eigenvalue}\right)  $ is
separated in two pieces\footnote{The analogue of Eq.$\left(  \ref{diff_eq}%
\right)  $ has been studied in the context of the stability of the
Schwarzschild solution for the first time by Regge and
Wheeler\cite{ReggeWheeler}, Vishveshwara\cite{Vishveshwara},
Zerilli\cite{Zerilli}, Press and Teukolsky, \cite{PressTeukolsky},
Stewart\cite{Stewart} and Chandrasekhar\cite{Chandrasekhar}. They concluded
that the Schwarzschild solution is \textit{classically} stable.\label{p1}}%
\begin{equation}
\frac{d^{2}H_{1}\left(  y\right)  }{dy^{2}}+b^{\pm}\left(  y\right)
\frac{dH_{1}\left(  y\right)  }{dy}+a^{\pm}\left(  y\right)  H_{1}\left(
y\right)  =-\mu^{2}H_{1}\left(  y\right)  , \label{diff_eq}%
\end{equation}
where%
\begin{equation}
b^{+}\left(  y\right)  =\sqrt{1-\frac{2\,MG}{r\left(  y\right)  }}%
\frac{\left(  4\,r^{2}\left(  y\right)  -23\,MGr\left(  y\right)
+27\,M^{2}G^{2}\right)  }{r^{2}\left(  y\right)  \left(  r\left(  y\right)
-2\,MG\right)  \left(  r\left(  y\right)  -3\,MG\right)  }\qquad a^{+}\left(
y\right)  =\frac{8MG}{\left(  r\left(  y\right)  -3\,MG\right)  r^{2}\left(
y\right)  }%
\end{equation}
in $\mathcal{\bar{M}}_{+}$ and%
\begin{equation}
b^{-}\left(  y\right)  =-\frac{\left(  4\,r^{2}\left(  y\right)
-23\,MGr\left(  y\right)  +27\,M^{2}G^{2}\right)  }{r^{2}\left(  y\right)
\left(  r\left(  y\right)  -3\,MG\right)  \sqrt{1-2\,MG/r\left(  y\right)  }%
}\qquad a^{-}\left(  y\right)  =\frac{8MG}{\left(  r\left(  y\right)
-3\,MG\right)  r^{2}\left(  y\right)  }%
\end{equation}
in $\mathcal{\bar{M}}_{-}$. After having cast Eq.$\left(  \ref{diff_eq}%
\right)  $ in the standard Sturm-Liouville form, we apply a variational
procedure to establish the existence of an instability by means of trial
functions. The standard Sturm-Liouville form of Eq.$\left(  \ref{diff_eq}%
\right)  $ is%
\begin{equation}
\frac{d}{dy}\left(  p^{\pm}\left(  y\right)  \frac{d}{dy}H_{1}\left(
y\right)  \right)  +q^{\pm}\left(  y\right)  H_{1}\left(  y\right)  +\mu
^{2}p^{\pm}\left(  y\right)  H_{1}\left(  y\right)  =0, \label{SturmLiouville}%
\end{equation}
where the integrating factor $p^{\pm}\left(  y\right)  $ is defined by%
\begin{equation}
p^{\pm}\left(  y\right)  =\exp\int b^{\pm}\left(  y\right)  dy=\exp\pm
\int\frac{b^{\pm}\left(  r\left(  y\right)  \right)  }{\sqrt{1-2\,MG/r\left(
y\right)  }}dr=\frac{\left(  r\left(  y\right)  \right)  ^{9/2}\left(
r\left(  y\right)  -2\,MG\right)  ^{3/2}}{\left(  r\left(  y\right)
-3\,MG\right)  ^{2}}. \label{intfac}%
\end{equation}
To compute the integrating factor we used Eq.$\left(  \ref{propdist}\right)
$. The coefficient
\begin{equation}
q^{\pm}\left(  y\right)  =p^{\pm}\left(  y\right)  a^{\pm}\left(  y\right)
\end{equation}
and the boundary conditions%
\begin{equation}
\left.  p^{+}\left(  y\right)  \frac{dH_{1}\left(  y\right)  }{dy}H_{1}\left(
y\right)  \right\vert _{y=0}^{y=+\infty}\qquad\text{and\qquad}\left.
p^{-}\left(  y\right)  \frac{dH_{1}\left(  y\right)  }{dy}H_{1}\left(
y\right)  \right\vert _{y=-\infty}^{y=0},
\end{equation}
lead us to write the Sturm-Liouville problem in the following functional form%
\begin{equation}
F\left[  H_{1}\left(  y\right)  ;\lambda\right]  =\frac{N\left[  H_{1}\left(
y\right)  ;\lambda\right]  }{D\left[  H_{1}\left(  y\right)  ;\lambda\right]
}=\mu_{\pm}^{2}\left(  \lambda\right)  =\frac{-\int_{I_{\pm}}dyp^{\pm}\left(
y\right)  \left(  \frac{d}{dy}H_{1}\left(  y\right)  \right)  ^{2}%
+\int_{I_{\pm}}dyq^{\pm}\left(  y\right)  H_{1}^{2}\left(  y\right)  }%
{\int_{I_{\pm}}dyp^{\pm}\left(  y\right)  H_{1}^{2}\left(  y\right)  },
\label{Var}%
\end{equation}
where $I_{+}\equiv\left[  0,+\infty\right)  $ and $I_{-}\equiv\left(
-\infty,0\right]  $. The choice of a trial function is suggested by the
asymptotic behavior of in Eq.$\left(  \ref{diff_eq}\right)  $. Indeed, when
$y\rightarrow\pm\infty,$ $H_{1}^{\pm}\left(  y\right)  \simeq\exp\left(
\mp\lambda y\right)  \simeq\exp\left(  -\lambda r\right)  $. The denominator
in Eq.$\left(  \ref{Var}\right)  $ compensates the lack of normalization in
the trial function $H_{1}^{\pm}\left(  y\right)  $. If $\lambda$ is defined as
a variational parameter, then the Rayleigh-Ritz method leads to the following
result\footnote{Details of computation can be found in the Appendix.}%
\begin{equation}
\mu_{\pm}^{2}\left(  \lambda_{\min}\right)  =-.1374637652\qquad for\qquad
\lambda_{\min}=.4471243737.
\end{equation}
However the denominator of Eq.$\left(  \ref{Var}\right)  $ vanishes for
$\bar{\lambda}=1.360771766$ and for $\lambda>\bar{\lambda}$, the
\textquotedblleft\textit{normalization}\textquotedblright\ changes sign. The
lack of positivity for every $\lambda$ is related to the appearance of a
spurious singularity in $r=3MG$\footnote{See details in Appendix \ref{app}}.
Therefore, it is necessary to choose a better form for the trial function
$H_{1}^{\pm}\left(  y\right)  $ which partially removes the singularity in
$r=3MG$. Such a form is given by $H_{1}^{\pm}\left(  y\right)  \simeq
H_{1}\left(  r\right)  =\exp\left(  -\lambda r\right)  \left(  r-3\,MG\right)
$. With this choice, the \textquotedblleft\textit{normalization}%
\textquotedblright\ is strictly positive for every $\lambda$. The minimum of
$\mu_{\pm}^{2}\left(  \lambda\right)  $ is reached for $\lambda_{\min
}=.9223711858$ and $\mu_{\pm}^{2}\left(  \lambda_{\min}\right)  =-.3827139389$%
. It is interesting to note that%
\begin{equation}
\mu^{2}\left(  \lambda_{\min}\right)  =\mu_{+}^{2}\left(  \lambda_{\min
}\right)  +\mu_{-}^{2}\left(  \lambda_{\min}\right)  =-0.765\,44\rightarrow
\mu^{2}\left(  \lambda_{\min}\right)  /\left(  2MG\right)  ^{2}%
=-0.191\,36/\left(  MG\right)  ^{2}.
\end{equation}
This is the same value obtained by GPY for a single Euclidean black hole.

\section{Summary and Conclusions}

\label{p5}In this paper we have analyzed, the one loop contribution coming
from quantum fluctuations around a background metric describing a
Schwarzschild wormhole which has boundary terms in both wedges of the Penrose
diagram. Usually, this kind of analysis is performed only in the $R_{+}$ wedge
of diagram $\left(  \ref{penrose}\right)  $. However, due to the symmetry
property of transformation $\left(  \ref{discrete}\right)  $, a contribution
from region $R_{-}$ is not unexpected. Moreover, the possibility of a fine
tuning between boundaries in the opposite regions gives the opportunity to
vanish the classical gravitational action. The vanishing of the classical
action leads to the interpretation that we are dealing with flat space. On the
other hand the presence in each wedge of a term proportional to the ADM mass
is a signal that a non flat configuration has been taken under examination. An
interpretation of this puzzling situation can be given in terms of black hole
pair creation mediated by a wormhole, with the pair elements residing in the
different universes. This proposal is not completely new and it has been
investigated in Refs.\cite{Remo} in a Hamiltonian formulation, where a
foliation of the manifold $\mathcal{M}$ is crucial. On the other hand, the
action formalism has the advantage of being fully covariant. However, the
vanishing of the classical action is not sufficient to establish if spacetime
will produce a pair or will persist in its flat configuration. It is,
therefore necessary to discover if the one loop approximation will produce an
instability. The variational approach we have used for the Sturm-Liouville
problem in the TT sector has shown an unstable mode in the s-wave
approximation in both wedges. This is the signal that flat space is not only
unstable when the space is furnished with a temperature $T=1/\left(  8\pi
MG\right)  $, but is even unstable with respect to Schwinger pair creation.
One possible conclusion is that flat space can be no more considered as the
general accepted vacuum of quantum gravity and the instability with respect to
Schwinger pair suggests to search in this direction instead of rejecting the
result with the further motivation that this instability can disappear if the
bounding box is restricted enough. A partial success in this direction has
been obtained by Allen\cite{Allen} in terms of hot quantum gravity from one
side. On the other side, the choice of a non-trivial vacuum formed by a large
number of black hole pair generated in the same way we have investigated in
this paper has led to a possible stable picture of \textit{space time foam},
which could replace flat space as a candidate for the quantum gravity
vacuum\cite{Remo1}.

\section{Acknowledgments}

I wish to thank M. Cadoni, R. Casadio, M. Cavagli\`{a}, V. Moretti and S.
Mignemi for useful comments and discussions.

\appendix

\section{Searching for Eigenvalues with the Rayleigh-Ritz method}

\label{app}In this Appendix, we will explicitly solve the eigenvalue problem
of Eq.$\left(  \ref{Var}\right)  $ with the Rayleigh-Ritz method. We here
report two principal choices of trial wave functions:

\begin{enumerate}
\item $H_{1}^{\pm}\left(  y\right)  =\exp\left(  \mp\lambda y\right)
\rightarrow\exp\left(  -\lambda r\right)  $. For this choice Eq.$\left(
\ref{Var}\right)  $ becomes,%
\begin{equation}
\mu^{2}\left(  \lambda\right)  =\lambda^{2}-8MG\frac{\int_{2MG}^{+\infty
}dre^{-2\lambda r}\frac{r^{3}\left(  r-2MG\right)  }{\left(  r-3MG\right)
^{3}}}{\int_{2MG}^{+\infty}dre^{-2\lambda r}\frac{r^{5}\left(  r-2MG\right)
}{\left(  r-3MG\right)  ^{2}}}=\lambda^{2}-\frac{A\left(  \lambda\right)
}{B\left(  \lambda\right)  },
\end{equation}
where%
\[
A\left(  \lambda\right)  =8MG\int_{2MG}^{+\infty}dre^{-2\lambda r}\frac
{r^{3}\left(  r-2MG\right)  }{\left(  r-3MG\right)  ^{3}}=256\left(
MG\right)  ^{3}\int_{1}^{+\infty}d\rho\exp\left(  -2\tilde{\lambda}%
\rho\right)  \frac{\rho^{3}\left(  \rho-1\right)  }{\left(  2\rho-3\right)
^{3}}%
\]%
\begin{equation}
=256\left(  MG\right)  ^{3}\left[  I_{3}^{4}\left(  \tilde{\lambda}\right)
-I_{3}^{3}\left(  \tilde{\lambda}\right)  \right]  \label{Alambda}%
\end{equation}
and%
\[
B\left(  \lambda\right)  =\int_{2MG}^{+\infty}dre^{-2\lambda r}\frac
{r^{5}\left(  r-2MG\right)  }{\left(  r-3MG\right)  ^{2}}=128\left(
MG\right)  ^{5}\int_{1}^{+\infty}d\rho\exp\left(  -2\tilde{\lambda}%
\rho\right)  \frac{\rho^{5}\left(  \rho-1\right)  }{\left(  2\rho-3\right)
^{2}}%
\]%
\begin{subequations}
\begin{equation}
=128\left(  MG\right)  ^{5}\left[  I_{2}^{6}\left(  \tilde{\lambda}\right)
-I_{2}^{5}\left(  \tilde{\lambda}\right)  \right]  . \label{Blambda}%
\end{equation}
We have defined $\rho=r/\left(  2MG\right)  $, $\tilde{\lambda}=\lambda2MG$
and introduced the following function
\end{subequations}
\begin{equation}
I_{m}^{n}\left(  \tilde{\lambda}\right)  =\int_{1}^{+\infty}d\rho\exp\left(
-2\tilde{\lambda}\rho\right)  \frac{\rho^{n}}{\left(  2\rho-3\right)  ^{m}%
}=\left(  -1\right)  ^{n}\frac{d^{n}I_{m}^{0}\left(  \tilde{\lambda}\right)
}{d\tilde{\lambda}^{n}}.
\end{equation}
$I_{m}^{0}\left(  \tilde{\lambda}\right)  $ can be easily integrated with the
formula\cite{GR}%
\begin{equation}
\int_{u}^{+\infty}dx\frac{\exp\left(  -\nu x\right)  }{\left(  x+\beta\right)
^{m}}=\exp\left(  -\nu u\right)  \sum_{k=1}^{m-1}\frac{\left(  k-1\right)
!}{\left(  m-1\right)  !}\frac{\left(  -\nu\right)  ^{m-k-1}}{\left(
u+\beta\right)  ^{k}}-\frac{\left(  -\nu\right)  ^{m-1}}{\left(  m-1\right)
!}\exp\left(  \beta\nu\right)  \operatorname{Ei}\left[  -\left(
u+\beta\right)  \nu\right]  \label{Formula}%
\end{equation}
with $m\geq2$, $\left\vert \arg\left(  u+\beta\right)  \right\vert <\pi$ and
$\operatorname{Re}\nu>0$. $\operatorname{Ei}\left(  x\right)  $ is the
exponential function. We get $A\left(  \lambda\right)  =$%
\begin{equation}
{\frac{\left(  -18\,{\lambda}\,{e^{{\lambda}}}-108\,{\lambda}^{3}%
\operatorname{Ei}\left(  {\lambda}\right)  +72\,{\lambda}^{2}\operatorname{Ei}%
\left(  {\lambda}\right)  +81\,{\lambda}^{2}{e^{{\lambda}}}-27\,{\lambda}%
^{3}{e^{{\lambda}}}+27\,{\lambda}^{4}\operatorname{Ei}\left(  {\lambda
}\right)  -2\,{e^{{\lambda}}}\right)  {e^{-3\,{\lambda}}}}{8{\lambda}^{2}}}%
\end{equation}
and $B\left(  \lambda\right)  =$%
\begin{equation}
{\frac{\left(  24\,{e^{{\lambda}}}-243\,{\lambda}^{5}{e^{{\lambda}}%
}+243\,{\lambda}^{6}\operatorname{Ei}\left(  {\lambda}\right)  -648\,{\lambda
}^{5}\operatorname{Ei}\left(  {\lambda}\right)  +194\,{\lambda}^{3}%
{e^{{\lambda}}}+72\,{\lambda}\,{e^{{\lambda}}}+126\,{\lambda}^{2}{e^{{\lambda
}}}+405\,{\lambda}^{4}{e^{{\lambda}}}\right)  {e^{-3\,{\lambda}}}}%
{128{\lambda}^{5}}.}%
\end{equation}
$B\left(  \lambda\right)  $ vanishes for $\lambda=1.360771766$ and $\mu
^{2}\left(  \lambda\right)  $ gets its minimum for $\lambda_{\min
}=.4471243737$ leading to $\mu^{2}\left(  \lambda_{\min}\right)
=-.1374637652$

\item With the help of Eq.$\left(  \ref{Formula}\right)  $ and choosing
$H_{1}^{\pm}\left(  y\right)  =\exp\left(  \mp\lambda y\right)  \left(
r\left(  y\right)  -3MG\right)  \rightarrow\exp\left(  -\lambda r\right)
\left(  r-3MG\right)  $ we get a more complicated form in $N\left[
H_{1}\left(  y\right)  ;\lambda\right]  $ of Eq.$\left(  \ref{Var}\right)  $,
while $D\left[  H_{1}\left(  y\right)  ;\lambda\right]  $ is simpler. Indeed%
\[
N\left[  H_{1}\left(  y\right)  ;\lambda\right]  =-\int_{2MG}^{+\infty}%
dr\frac{r^{5}\left(  r-2\,MG\right)  }{\left(  r-3\,MG\right)  ^{2}}\left(
\frac{d}{dr}\left[  e^{-2\lambda r}\left(  r-3MG\right)  \right]  \right)
^{2}+8MG\int_{2MG}^{+\infty}dre^{-2\lambda r}\frac{r^{3}\left(  r-2MG\right)
}{\left(  r-3MG\right)  }%
\]%
\[
=-\frac{\left(  2MG\right)  ^{7}}{\left(  2MG\right)  ^{2}}\int_{1}^{+\infty
}d\rho\frac{\rho^{5}\left(  \rho-1\right)  }{\left(  2\rho-3\right)  ^{2}%
}\left(  \frac{d}{d\rho}\left[  e^{-2\tilde{\lambda}\rho}\left(
2\rho-3\right)  \right]  \right)  ^{2}+8MG\frac{\left(  2MG\right)  ^{5}}%
{MG}\int_{1}^{+\infty}d\rho e^{-2\tilde{\lambda}\rho}\frac{\rho^{3}\left(
\rho-1\right)  }{\left(  2\rho-3\right)  }%
\]%
\begin{equation}
=-\,\left(  2MG\right)  ^{5}{\frac{\left(  243\,{\lambda}^{6}{Ei}\left(
{\lambda}\right)  -239\,{\lambda}^{5}{e^{{\lambda}}}+42\,{e^{{\lambda}}%
}+225\,{\lambda}^{4}{e^{{\lambda}}}-432\,{\lambda}^{5}{Ei}\left(  {\lambda
}\right)  +86\,{\lambda}^{2}{e^{{\lambda}}}+102\,{\lambda}^{3}{e^{{\lambda}}%
}+78\,{\lambda}\,{e^{{\lambda}}}\right)  {e^{-3\,{\lambda}}}}{16{\lambda}^{5}%
}}%
\end{equation}
and%
\[
D\left[  H_{1}\left(  y\right)  ;\lambda\right]  =\int_{2MG}^{+\infty
}dr\left[  e^{-2\lambda r}r^{5}\left(  r-2\,MG\right)  \right]  =\left(
2MG\right)  ^{7}\int_{1}^{+\infty}d\rho\exp\left(  -2\tilde{\lambda}%
\rho\right)  \rho^{5}\left(  \rho-1\right)
\]%
\begin{equation}
=\left(  2MG\right)  ^{7}{\frac{{e^{-2{\lambda}}}\left(  45+2\,{\lambda}%
^{5}+10\,{\lambda}^{4}+30\,{\lambda}^{3}+60\,{\lambda}^{2}+75\,{\lambda
}\right)  }{8{\lambda}^{7}}.}%
\end{equation}
Note the positiveness of $D\left[  H_{1}\left(  y\right)  ;\lambda\right]  $
for every $\lambda$, ensuring a correct normalization. The minimum of $\mu
^{2}\left(  \lambda\right)  $ is reached in this case for $\lambda_{\min
}=.9223711858$ and $\mu^{2}\left(  \lambda_{\min}\right)  =-.3827139389$.
\end{enumerate}

\end{document}